# CenterYou: A cloud-based Approach to Simplify Android Privacy Management

Seyedmostafa Safavi*, Zarina Shukur

*Unit of Cyber Security, Faculty of Information Science and Technology, Universiti Kebangsaan Malaysia,43600 Bangi, Malaysia*

## ABSTRACT

With mobile applications and associated services becoming increasingly popular, concerns are being raised about security vulnerabilities and private data leakages of Smartphone users. Previous solutions to this well-known set of problems have approached it from the ground up, whereby the focus was on implementing reasonable security policies within Android's open source kernel. While these solutions have achieved the goals of improving Android, the way in which the polices have been implemented required re-writing the operating system source code, which is unnecessary and burdensome. In this work, a framework referred to as CenterYou is proposed to overcome these issues. It applies pseudo data technique and cloud-based decision-making system to scan and protect Smartphone devices from unnecessarily requested permissions by installed applications and identifies potential privacy leakages. The current paper demonstrated all aspects of the CenterYou application design, as well as described how the CenterYou framework was evaluated through a three-step evaluation process. To evaluate effectiveness of the design a case study was conducted by using a test-bed that was applied to both standard device, and that equipped with CenterYou. Efficiency evaluation was conducted, with the help of three popular benchmark applications (AnTuTu, Softweg, and Elixir 2) to assess the performance of the CenterYou application in managing the privacy in Android devices. Finally, in the third stage, a pilot study was conducted to show degree of usability evaluation involving seven participants. The findings indicated that CenterYou met its design objectives, as 68% of the participants were satisfied with its features. With the help of the Dalvik hooking method and cloud processing, the application was successful in making negligible contribution to the CPU usage, with 0.01 percent.

* E-mail: Safavi@takhosting.info

## 1. INTRODUCTION

Owing to the growing prevalence of smart gadgets, issues pertaining to protection and privacy are emerging. Most Smartphone users install applications on their devices in order to obtain additional features, such as games, or gain access to Internet sites, such as Facebook. Applications are sometimes over-privileged, as many require access to resources they do not need to function. Owing to this feature, such applications increase the impact of vulnerabilities and exposure to risk.

For example, an increasing number of consumers are using Smartphones for mobile banking and sharing other sensitive information like health information (Safavi & Shukur 2014), thus opening up possibilities for unauthorized access to their data. A further issue pertains to the computer storage and processing of classified information, which also present significant security and privacy risks. In 1999, Sun Microsystems chief executive officer Scott McNealy made a prophetic judgment of online privacy, stating "You have zero privacy anyway." In simple terms, privacy might be defined as the power of individuals to choose when, how and what type of data about them is revealed to others. In sum, privacy principles (Fischer-Hübner 2001); (Federrath 2001) require that systems minimize personalized data accumulation by, for instance, data anonymization.

Android's permission system is rather rigid and lacks flexibility. Users can only install applications by granting all permissions requested by that application. It is not possible to withhold specific permissions, either during installation or after the installation process. Thus, if the users are concerned about security, their only option is to uninstall an application. In other words, the current permission-based control is completely static. In addition, once an application is installed, users can neither see the resources the application accesses, nor they can permit or deny any such access. Once again, all they can do if they wish to revoke permissions is to uninstall the application from the device.

In order to mitigate these issues, many researchers have attempted to find practical means of enhancing the safety measures in the Android OS. For example, (Enck et al. 2008) proposed a light and portable certification software program based on analyzing the particular mixture of permissions pertinent to most popular Android software applications. In a more recent study, (Shin et al. 2011) presented a permission system for the Android platform, by identifying exactly how permissions, apps, factors, authorizations and states operate. In order to increase user security, (Nauman & Khan 2011) presented Apex, the policy enforcement platform intended for Android, that allows end-users to be able to selectively grant permissions to software applications, as well as enforce restrictions upon use of device resources. (Ongtang et al. 2012) presented an improved design of platform facilities that manage install-time permission tasks. While these are only a few examples of research initiatives intended for improving consumer safety, in recognition of growing issues imposed by malicious applications, several methods are already used to investigate the potential threats and mitigate them wherever possible. For example, (Noel et al. 2009) described a technique that evaluates weakness dependencies and displays most probable attack routes over a network. In another study, using permission-based protection model pertaining to the actual Android operating system, (Barrera et al. 2010) conducted a great empirical evaluation of security threats. Their findings highlighted the significance of protecting classified personalized information in Android portable devices.

The problem this paper aims to address is allowing users to manage application permissions without the need for excessive technical knowledge (Stanton et al. 2005). In addition, the goal is to allow users to spend less time responding to warning messages that ask for permission to access resources in device (Maximilien et al. 2001); (Leslie et al. 2005). It is expected that having proper protection would make Smartphone users more satisfied with the device.

The aim of this study is to design and implement security architecture for the Android operating system that would address the present challenges. This can be done by designing a user-friendly framework that satisfies the needs of non-technical users, implementing architecture that will run efficiently with fewer delays and develop the system that can effectively protect user's personal data by using a pseudo technique.

## 2. CENTERYOU: REDESIGN PERMISSIONS FOR REAL-TIME PRIVACY PROTECTION WITH HELP OF CLOUD

This section presents the methodology adopted in conducting this research—design science research (DSR) (Peffers et al. 2007). This particular design is inspired by the need to enhance the environment through the creation of novel and revolutionary artifacts, along with the techniques for constructing these types of artifacts. It consists of five phases, which explained here:

### 2.1 Problem Identification And Motivation Phase

In first phase, research problem is identified, and a literature research conducted following an extensive review of literature on mobile technology, privacy and current Android mechanisms. In addition, the relevance of the intended study is evaluated, often through interviews with experts in the field.

The problem this paper aims to address is allowing users to manage application permissions without the need for excessive technical knowledge (Stanton et al. 2005). In addition, the goal is to allow users to spend less time responding to warning messages that ask for permission to access resources in device (Maximilien et al. 2001); (Leslie et al. 2005). It is expected that having proper protection would make Smartphone users more satisfied with the device.

To protect users' personal information from over-privileged apps, a new mode of privacy is needed in Smartphones, whereby the access to user's personal information is controlled either by the user, or an automated process managed by cloud service. Furthermore, the user should have run-time control to modify the previously given permission.

### 2.2 Objectives Of Solution PHASE

The second phase requires the researcher to state the study objectives, based on the findings yielded by the literature review. As the issues affecting the current mechanism have been identified, solutions to the aforementioned problems can be discussed. This study aimed to design and implement security architecture for the Android operating system that would address the present security challenges. This has been achieved by designing a security framework, which would serve as an appropriate ecosystem for different security and privacy-protecting models. This research thus focused on usability, efficiency and effectiveness (Guler et al. 2002; Zhang & Adipat 2005; Kenteris et al. 2009; Coursaris & Kim 2011; Safavi et al. 2013) of the framework design, aiming to provide the ability to scan and protect the devices from unnecessarily requested permissions by installed applications. In line with these objectives, the artifact is also designed in this phase. This helps identify the aims that the artifact is expected to accomplish.

### 2.3 Design And Development Phase

The CenterYou framework modifies the Package Manager, whereby the set of permissions in system/centeryou folder is duplicated to ensure that every single authorization possesses both the 'Real' and a 'Pseudo' type. In case of pseudo permission, the API call provides a fake result to the application. To provide pseudo data, CenterYou application uses Zygote to enable the injection service. Once the framework installation is complete, an extended app_process executable is copied to /system/bin. This extended startup process adds an additional jar to the classpath and calls the methods contained within.

One of the differences between CenterYou and other solutions discussed as a part of the related works is that the former benefits from the cloud-based support system. This feature allows the user to keep the application in Autopilot (easy mode) and thus not have to worry about protecting his/her privacy from newly installed applications. One of the best examples of this service is antivirus software that has been in use for a long time. Still, the key difference between the CenterYou framework and any antivirus software currently on the market is the service they are supposed to provide to their respective users.

The cloud service is responsible for handling and managing all application permissions lists, sending notifications (settings and one-way secure messaging) to a particular user, and having space for backup of permissions list sent by each device application. To perform the aforementioned functions, this researcher used Amazon Web Services (AWS) (Services 2014) for the cloud service and PHP and Java programming languages to design and develop the website and notification system under the GCM service (Android 2014).

## 2.4 Demonstration And Evaluation Phase

The demonstration and evaluation of the designed framework is included in the fourth phase.

### a. Demonstration of CenterYou

This part includes autopilot/advance menu, application setting, apply permission setting through cloud helpline, setting and timer section, CenterYou as a cloud-based decision-making and monitoring system, and the CenterYou cloud notification service.

CenterYou starts with a splash page, shown in Figure 1 and after displaying this initial screen for three seconds, the splash page will be replaced by the main section of the CenterYou application.

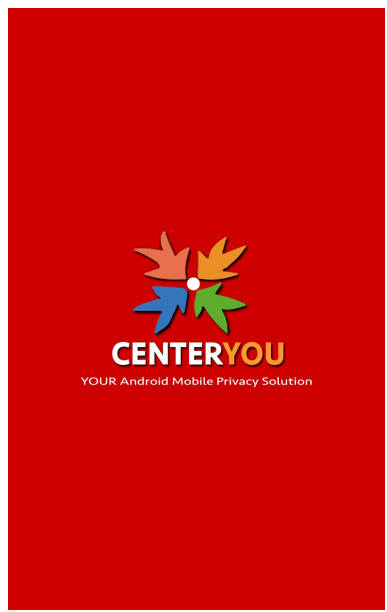

Figure 1: Splash Page for CenterYou

In this main page, the user has three options to select from, along with the calculation score pertaining to the Smartphone privacy. The options provided on this page allow the user to learn about the applications that are moderated by CenterYou, as well as see those that pose greater risk. As shown in Figure 2 and 3, the calculation allows the user to appreciate the privacy protection at the moment of usage, via the numerical score and usage of different colors. To start modifying or to allow the application to run in the autopilot mode, the user should choose the easy/advance option, as shown in Figure 3.

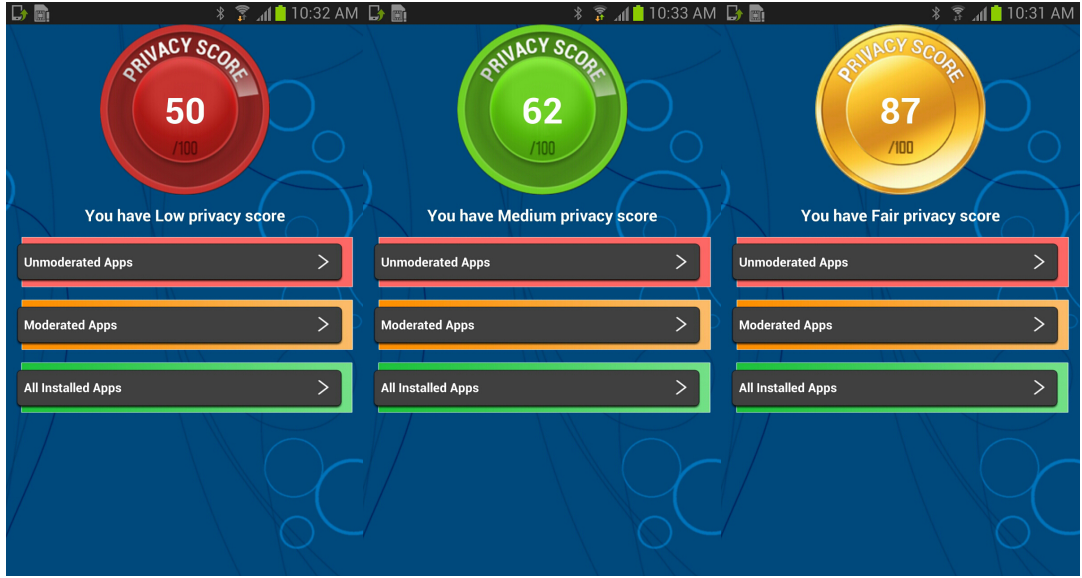

Figure 2: The privacy score on CenterYou

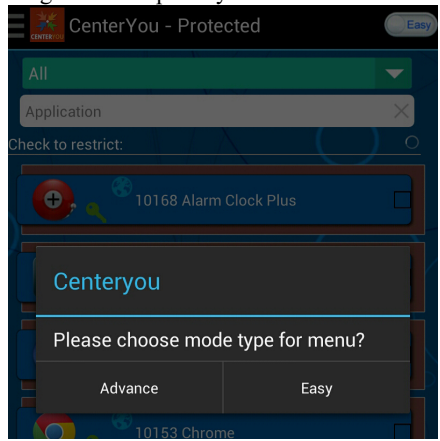

Figure 3: Popup to choose the method of management system in the CenterYou App

After selecting the last option, "All installed Apps", the user will be presented with a new screen, shown in Figure 3. If this is the first time the CenterYou application is used on this device, an option will pop up, prompting the user to choose one of the alternatives provided. This option allows the user to select between running the application on autopilot, or in manual mode. The user may change the selected option at any time, by touching the icon in the right top corner of the screen in this page. In this demonstration, "easy" option is chosen by default, but it can be changed at any time.

As shown in Figure 4, the user is given limited number of options, and most of the decisions are taken by the cloud. These features include backup, import and export, cloud help support system, and upload options, as well as timer to set a time to upload the information from the device to the cloud.

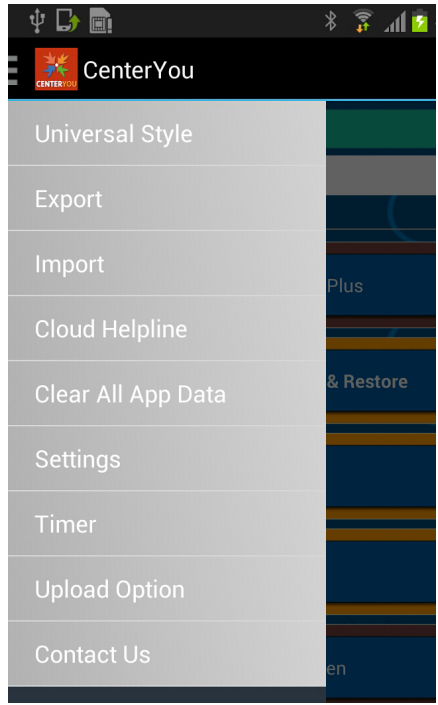

Figure 4: Menu designed for Autopilot or Easy Mode

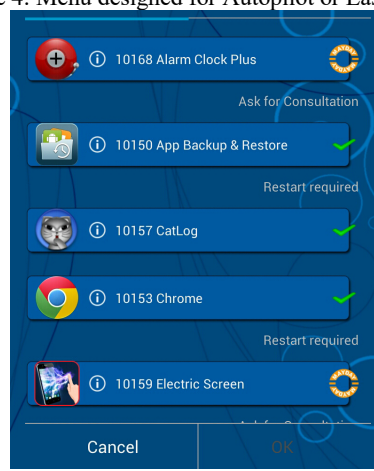

Figure 5: Applying permissions to all applications available on a specific device

As shown in Figure 5, CenterYou is fetching data from the cloud and applying it to the application in real time. This process takes four seconds per application. When the application is not available in the cloud database, the user can select the "Mayday" option to ask for a consultation. At this point, the application will send the file to the cloud monitoring system. After the moderation, the cloud will respond to the consolation requester with auto path apply, thus ensuring that the user will never need to do anything in this context. After receiving the notification, the user can be confident that the application can be used with no fear of privacy leakage.

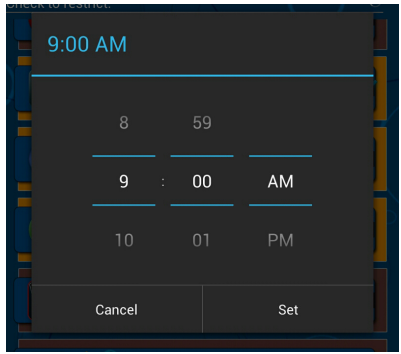

Figure 6: Timer for backup transmission

If the user has selected the "Autopilot" mode, the default time for backup is set to 9 AM in Figure 6. At this time, each morning, all data on the device is backed up automatically to the cloud server. However, the users can also be certain that the application is mindful about the price of data transmission, as many devices are working on a data plan.

As shown in Figure 7, the daily scheduled backup may not take place, if the device is in the WIFI only mode, or is using data plan (GPRS) at the time of the backup.

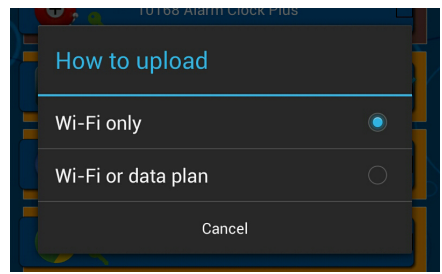

Figure 7: WIFI only or data plan option

The solution objectives will be evaluated based on the output from the evaluation part. Here, three approaches are adopted to help evaluate the framework and proof of solution objectives, as discussed in the following sections.

The Cloud is background of the processes employed to protect the user data is discussed, with the specific focus on the differences between CenterYou and other available frameworks. This section is designed to help administrators to remotely control every aspect of security and privacy in a targeted device. In other words, after the first installation of CenterYou on a targeted device, the cloud monitoring system will alert the administrator to control its operation when needed. In the first stage, the IMEI number of the device will be recorded and will be subsequently used to identify the correct device owner.

In the second stage, if the user keeps the device in the autopilot (easy) mode, the application will automatically ask the cloud to apply all proper application permissions lists to the device. In this stage, if any of the applications on user's device are not in the cloud list, the user will be sent a notification. In return, the administrator will receive an application package file (APK file), and will perform appropriate actions in order to resolve the issue of the unknown application.

More specifically, for each unknown application, the administrator will send the file for monitoring. Once the monitoring step is complete, the permissions will be saved in the cloud and applied on the device that requested the new permissions list via the original consultation.

Finally, the notification will be shown to the user, informing him/her that the application is ready and safe to use. Figure 8 displays a part of the monitoring system.

Figure 8: The application request page

### b. Evaluation of CenterYou

To demonstrate the effectiveness when evaluating of the research solution discussed in this paper, the researcher built a proof of concept process named "CenterYou", and conducted several user and case studies. This evaluation helped to prove the solution objective number three "To develop the system that can effectively protect user's personal data by using a pseudo technique." The evaluation discussion is presented in three parts, pertaining to the case study, performance evaluation, and a user study, respectively. These three evaluations pertain to the full scope of this study, which is subsequently discussed as well.

- **Effectiveness Evaluation**

The first step performed when evaluating a framework is a case study, as it can help to explore the application responses in different situations. In addition, it can serve as a proof of the ability of the design to meet the requirements, as well as assist the researcher in establishing the limits of the framework's ability, based on its response to test-bed attacks.

In this study, Google Chrome Browser and Simple Flash Light (designed by the researcher to attack the application) have been utilized and the results and responses from the application have been subsequently discussed. After applying the CenterYou restrictions to the application, the user accesses a regular website, such as maps.google.com. This example is used here, to show the user's current location. As can be seen in Figure 9, Google Maps cannot differentiate between the GPS location and the results manufactured by the pseudo. Thus, it presents the fake location as a current map location of the user.

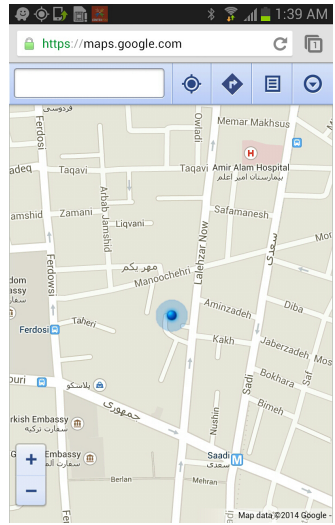

Figure 9 CenterYou Chrome browser response

This practical example clearly confirmed that it was successful in protecting user's location and secured it from the use by unauthorized applications. Simple Torch application is an application designed and developed to test-bed the framework in different contexts. This application is designed to be over-privileged, so that this case can be tested. As can be seen in Figure 10, the backend part of the system (the cloud server) has received the user data upon every execution of the application. At the same time, the user information, such as his/her location and the front face camera photo that was taken, was sent to the server. This process is a result of not having the protection of the CenterYou application installed on the device. Thus, the privacy of the mobile owner is at risk.

Figure 10 shows the case when the user has changed his/her location data manually via the CenterYou application. Now, this fake data applies to the Simple Torch application. Now, the application permissions list will be blocked and the user can set up a new location using the pseudo technique. This change will protect user privacy, as the same fake address will be sent to the server that manages the Simple Torch application, as shown in Figure 6.

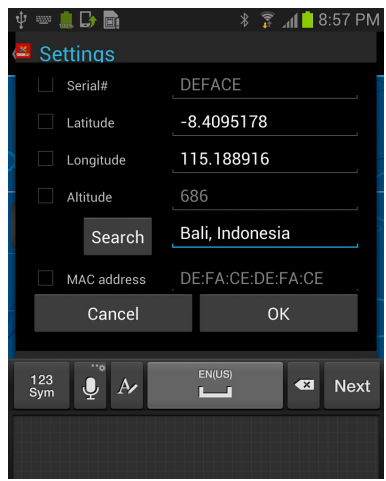

Figure 10 CenterYou settings UI used to manipulate location on Simple Torch

Figure 11 Simple Torch backend server

Figure 12 Simple Torch backend server after applying the pseudo technique to manipulate the location data

In Next, the user has manipulated the IMEI number by changing it to 123456 via CenterYou setting. As shown in Figure 13, the application responds to this request by sending the fake data to the Simple Torch server, thus protecting the user's device from unauthorized access to sensitive information.

Figure 13 Simple Torch backend server after applying the pseudo technique to manipulate data for IMEI and location

With the help of this case study, the effectiveness of the framework in terms of protecting user's personal data by using CenterYou application has been demonstrated. This also serves to fulfill every goal of the present research. This practical examination of the way the user's location and IMEI was protected by CenterYou from a test-bed attack and an over-privileged application thus confirms that it is effective in fulfilling all its design objectives.

- **Efficiency Evaluation**

In order to assess whether CenterYou met the solution objective number two "To implement architecture that will run efficiently with fewer delays," a further test was conducted. This is necessary, in order for the application to be accepted by the mainstream Android users. In other words, CenterYou must not impose significant runtime performance and mobile CPU usage overheads. Empirical evidence indicates that most users would be unwilling to give up functionality in order to gain protection and security (Nauman et al. 2010). Thus, the CenterYou application must not slow down the processor. In order to ascertain that this is not the case, a performance measurement was conducted, using three well-known benchmark applications found in the Google Play Store, namely AnTuTu (Antutu-labs 2014) by AnTuTu Labs, Elixir 2 by Tamás Barta (Barta 2014) and Benchmark by Softweg (Softweg 2014). The benchmark testing was executed without and with the CenterYou framework applied on the same device. A large number of trials were run in order to obtain statistically meaningful results. In each benchmark test, the same number of apps/services was loaded and running at any time, to ensure consistency and comparability of the results yielded.

## I. Physical Performance Measurement

AnTuTu benchmark, Softweg and Elixir 2 were used to evaluate the physical performance of the CenterYou application. The results of these benchmark tests provided the data that could ascertain whether the CenterYou application would affect the device performance. Ideally, this is achieved by not utilizing the physical elements because they have an indirect effect on the device performance. The results of the 120 runs that were performed using the AnTuTu benchmark on both original Android and that augmented by CenterYou are shown in Table 1.

Table 1: An TUTU comparative benchmark test results

|  | Original Device | | Applied CenterYou | |
| --- | --- | --- | --- | --- |
|  | Mean | Standard Deviation | Mean | Standard Deviation |
| CPU Integer | 630.75 | 129.41 | 611 | 27.86 |
| CPU Float Point | 626 | 34.91 | 635.5 | 9.53 |
| Ram Operation | 547 | 2.44 | 538.5 | 8.96 |
| 2D Score | 1136 | 59.63 | 1130.25 | 49.14 |
| 3D Score | 3188.5 | 10.63 | 3203.25 | 29.38 |
| Storage I/O | 405.25 | 49.08 | 425 | 16.79 |
| Database I/O | 427.5 | 35.70 | 397.5 | 45.55 |

The CPU integer and float tests clearly demonstrate that the device was unaffected by the CenterYou application, as the changes were minimal, from 630.75 to 611, as running CenterYou did not involve any system calls. The RAM operation test results, on the other hand, show a slight decline, from 547 to 538.5, since the application runs most of the tasks in the Dalvik VM process and thus never makes the RAM busy by sending commands. The 3D and 2D score tests, which are measurements of visual functionality, were also not affected, as only six changes were made through CenterYou. Thus, these tests confirm that CenterYou creates minimal overhead, with all differences in performance within one standard deviation of the initial Android outcome.

The outcome with regard to 120 runs with the Softweg benchmark, on both the original device and that with CenterYou installed, is shown in Table 2. The minimal change, from 22.60 to 21.67, confirms that the application did not place any burdens on the Android device. Graphics ratings were also unaffected by the installation and use of CenterYou. Once again, these tests confirm that CenterYou creates a minimal overhead, with all performance indicator scores within one standard deviation from those pertaining to the original results.

Table 2: Softweg comparative benchmark test results

|  | Original Device | | Applied CenterYou | |
| --- | --- | --- | --- | --- |
|  | Mean | Standard Deviation | Mean | Standard Deviation |
| Graphics Score | 22.60 | 1.41 | 21.67 | 3.54 |
| CPU Score | 3910.09 | 49.21 | 3730.44 | 153.62 |
| Memory Score | 746.65 | 24.83 | 729.56 | 37.84 |
| Create Files | 8.37 | 0.36 | 13.59 | 2.21 |
| Read File | 220.28 | 105.87 | 192.47 | 52.83 |
| Write File | 8.5 | 0.98 | 7.74 | 1.54 |
| Delete File | 22.51 | 1.68 | 22.12 | 2.89 |

In file system read/write tests, the actual rate (M/sec) was calculated, along with erase tests. The goal was to establish time (seconds) required to create or erase 1000 empty documents. The difference between the original device performance and that when CenterYou was installed was slight, from 220.28 to 192.47 and 8.5 to 7.74, for creation and erasing tasks respectively. These findings indicate that CenterYou uses fewer physical devices, confirming that the device owner will not notice any changes in the device processing time and speed upon installation of this application. This finding thus confirms that the application has fulfilled every aspect of this research in terms of its performance.

The outcome regarding 120 runs with the Elixir benchmark with both the original device and that augmented by CenterYou is shown in Table 3. The actual memory space, along with central processing unit results, was unaffected by the usage of the CenterYou software application, since it does not require any type of system calls. To conclude, as shown in Table 3, with the help of the Dalvik hooking method and cloud processing, the CenterYou application contributed only 0.01 percent to the Smartphone total CPU usage.

Table 3: Elixir 2 comparative benchmark test results

|  | Applied CenterYou | |
| --- | --- | --- |
|  | Mean | Standard Deviation |
| Mobile total CPU usage | 0.015 | 0.0076 |

## II.   Leakage Test Response

The next stage of the performance test required evaluating CenterYou application with respect to the protection it provided against data leakage (Figure 14). In this step, Elixir 2 was used to find out how effectively CenterYou can protect user's data from leakage to different applications. After applying the CenterYou settings option for Elixir 2, as shown in Figure 8, the pseudo technique was applied and tested by running the Elixir 2 application. The original system IMEI number was changed to 049359160684869 and the location (latitude, longitude) was set to Mashhad, Iran. In addition, the user's address was changed to NK, and the Mac address was modified to 74:E3:FE:76:2C:90, while the IP address was set to 0.0.0.0. Finally the data connection was modified to allow traffic, enabling the user to receive a report from Elixir 2 on this position.

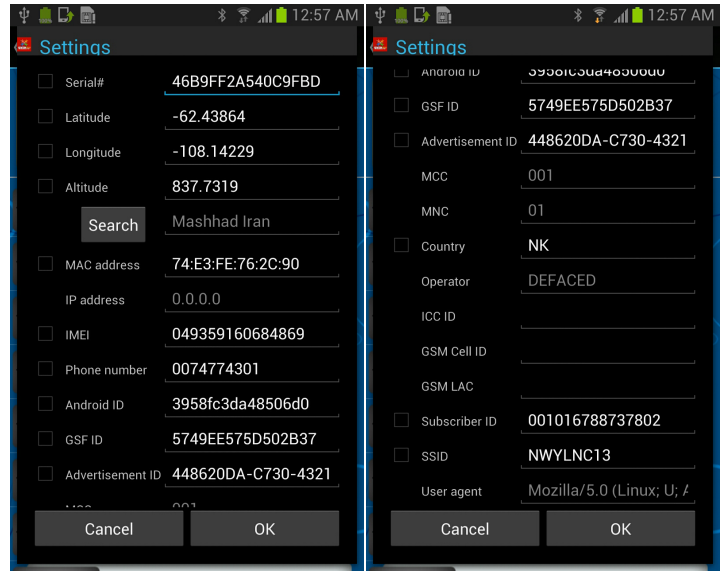

Figure 14: CenterYou application settings page for Elixir 2

- **Usability Evaluation**

This section provides the demonstration of the CenterYou performance through a user study. To assess the CenterYou application feasibility and efficacy, the researcher conducted interviews with seven individuals, who were purposefully selected, as likely end-users.

Prior to conducting individual interviews, the researcher conducted a brief lecture, during which all attendees learned about privacy and the goals of the CenterYou application. The goal of the short seminar was to encourage the participants to focus on the value of ideas that were implemented in the newly developed application.

Next, the researcher demonstrated how the CenterYou was used and explained its performance in practice. In the workshop, conducted as a single five-hour session, the participants first received manual instructions, after which they were instructed to test CenterYou in two environments. Upon completion of the evaluation, all participants were given two questionnaires, which they completed based on the experience and information they gained while working with both environments.

I.   **Usability test**

The objective of testing the CenterYou usability was to establish whether the design meets the expectations typical users. The discussion of the usability test is presented in two parts, corresponding to the tasks the study participants performed. First, the participants tested the device without the application, and subsequently assessed its performance after the first CenterYou prototype was installed.

The second stage of the pilot study, as noted above, involved seven participants. In order to gauge the utility of the CenterYou design, this researcher examined the participants' responses to the survey questions, and observed them during the tests. This information was subsequently used to improve the application performance.

The study participants completed the questionnaire (Table 4), and their individual responses with respect to their perceptions of CenterYou and Mockdroid (Beresford et al. 2011) privacy protection frameworks were used to assess the utility of the application. The main reason behind comparing CenterYou with Mockdroid was that the latter had already introduced a novel technique in faking data. In addition, its features pertaining to protecting privacy in Android environment are similar to those of CenterYou. The participants were given the opportunity to test both frameworks, and could thus provide views on the quality and utility of CenterYou relative to a similar application. Analysis of their answers assisted the researcher in determining the facets of this particular software application that need changes, as well as those that meet user requirements. The questionnaire used in this pilot study was previously designed and employed by Lewis, who conducted usability measurements at IBM (Lewis, 1993). However, in order to apply this instrument to the CenterYou and Mockdroid frameworks, it required some modifications. For user data collection, Likert method was chosen (Likert 1932), as it allows the respondents to choose a number corresponding to a specific statement that best reflects their level of agreement/disagreement with a particular questionnaire item.

The seven individuals that took part in this test were students or research scholars from various universities across Malaysia. The first requirement for participation in the study was that the participant previously owned Android mobile phone, and knows how to install and work with the device and applications. In addition, the participants were expected to have knowledge that would allow them to perform root and flash to different ROM on their own devices. The study participants came from UKM, UPM, and Apit University and the group included both males and females. To fulfill the objectives of this study, all usability study participants were non-technical mobile phone users.

Table 4: Usability questionnaire for CenterYou and Mockdroid

| Row number | Questions |
| --- | --- |
| 1. | The installation was easy to process |
| 2. | The user interface (UI) of this application was pleasant |
| 3. | The Easy/Advanced mode was useful (In the case of availability) |
| 4. | Overall, I am satisfied with how easy it can be to set up this application |
| 5. | I felt comfortable using the features of this framework |
| 6. | It was easy to learn to use this framework |
| 7. | The information (such as notification messages, easy/advanced option) provided with this application was clear (In the case of availability) |
| 8. | It was easy to work with the cloud when needed (In the case of availability) |
| 9. | The information was effective in helping me complete the tasks and scenarios |
| 10. | The cloud decision-making system was effective in most of the scenarios (In the case of availability) |
| 11. | The organization of information on the different user interfaces was clear |
| 12. | This application has all the functions and capabilities I would expect it to have |
| 13. | The autopilot (easy mode) was really useful in handling this application (In the case of availability) |
| 14. | Overall, I am satisfied with this application |
| 15. | I would recommend this application to others |

**Please note that the scale is anchored at 5 = "Strongly agree" and 1 = "Strongly disagree"**
**If a particular functionality is not available, leave it blank**

Table 5: Comparison of the Mockdroid and CenterYou framework trough a questionnaire

| | Question Number | CenterYou Application | | | | | | | Mockdroid Application | | | | | | |
|---|---|---|---|---|---|---|---|---|---|---|---|---|---|---|---|
| | | A1 | A2 | A3 | A4 | A5 | Arithmetic Mean | Standard Deviation | A1 | A2 | A3 | A4 | A5 | Arithmetic Mean | Standard Deviation |
| UI Quality Questions | Q2 | 0 | 0 | 0 | 4 | 3 | 4.43 | 0.53 | 0 | 2 | 3 | 2 | 0 | 3 | 0.82 |
| | Q3 | 0 | 0 | 0 | 4 | 3 | 4.43 | 0.53 | 0 | 0 | 0 | 0 | 0 | 0 | 0 |
| | Q7 | 0 | 0 | 0 | 4 | 3 | 4.43 | 0.53 | 0 | 0 | 0 | 0 | 0 | 0 | 0 |
| | Q9 | 0 | 0 | 0 | 3 | 4 | 4.57 | 0.53 | 3 | 1 | 3 | 0 | 0 | 2 | 1 |
| | Q11 | 0 | 0 | 0 | 3 | 4 | 4.57 | 0.53 | 1 | 1 | 4 | 1 | 0 | 2.71 | 0.95 |
| | Q13 | 0 | 0 | 0 | 0 | 7 | 5 | 0 | 0 | 0 | 0 | 0 | 0 | 0 | 0 |
| Design Goal Questions | Q1 | 0 | 0 | 0 | 2 | 5 | 4.71 | 0.49 | 3 | 3 | 1 | 0 | 0 | 1.71 | 0.76 |
| | Q4 | 0 | 0 | 0 | 2 | 5 | 4.71 | 0.49 | 4 | 2 | 0 | 1 | 0 | 1.71 | 1.11 |
| | Q5 | 0 | 0 | 0 | 5 | 2 | 4.29 | 0.49 | 0 | 3 | 3 | 1 | 0 | 2.71 | 0.76 |
| | Q6 | 0 | 0 | 0 | 3 | 4 | 4.57 | 0.53 | 1 | 2 | 4 | 0 | 0 | 2.43 | 0.79 |
| | Q8 | 0 | 0 | 0 | 4 | 3 | 4.43 | 0.53 | 0 | 0 | 0 | 0 | 0 | 0 | 0 |
| | Q10 | 0 | 0 | 0 | 1 | 6 | 4.86 | 0.38 | 0 | 0 | 0 | 0 | 0 | 0 | 0 |
| | Q12 | 0 | 0 | 0 | 2 | 5 | 4.71 | 0.49 | 0 | 5 | 2 | 0 | 0 | 2.29 | 0.49 |
| | Q14 | 0 | 0 | 0 | 2 | 5 | 4.71 | 0.49 | 3 | 0 | 4 | 0 | 0 | 2.14 | 1.07 |
| | Q15 | 0 | 0 | 0 | 0 | 7 | 5 | 0 | 4 | 2 | 1 | 0 | 0 | 1.57 | 0.79 |

As shown in Table 5, the participants' responses indicated that the CenterYou application was a significant improvement compared to Mockdroid.

The different results pertained to the satisfaction with using the cloud service response, the application itself, the changes in the Android operating system and making the system more user-friendly.

## II. Results pertaining to the change from Mockdroid to the CenterYou framework

Table 6 shows the improvements achieved by CenterYou, relative to Mockdroid, expressed as a percentage.

Table 6: Improvement percentage from Mockdroid to CenterYou framework

| Question Number | Mockdroid Mark | CenterYou Mark | Change | Percentage of Improvement |
|---|---|---|---|---|
| Question 1 | 1.71 | 4.71 | 3 | 63.69426752 |
| Question 2 | 3 | 4.43 | 1.43 | 32.27990971 |
| Question 3 | Not available | 4.43 | No VALUE! | No VALUE! |
| Question 4 | 1.71 | 4.71 | 3 | 63.69426752 |
| Question 5 | 2.71 | 4.29 | 1.58 | 36.82983683 |
| Question 6 | 2.43 | 4.57 | 2.14 | 46.82713348 |
| Question 7 | Not available | 4.43 | No VALUE! | No VALUE! |
| Question 8 | Not available | 4.43 | No VALUE! | No VALUE! |
| Question 9 | 2 | 4.57 | 2.57 | 56.23632385 |
| Question 10 | Not available | 4.86 | No VALUE! | No VALUE! |
| Question 11 | 2.71 | 4.57 | 1.86 | 40.70021882 |
| Question 12 | 2.29 | 4.71 | 2.42 | 51.38004246 |
| Question 13 | Not available | 5 | No VALUE! | No VALUE! |
| Question 14 | 2.14 | 4.71 | 2.57 | 54.56475584 |
| Question15 | 1.57 | 5 | 3.43 | 68.6 |

## 2.5 communication phase

In the fifth phase, referred to as communication, the results are published in different scholarly publications, thus helping disseminate the innovating ideas generated through the research and complete the study.

## 3. RELATED WORK

To address the growing prevalence of spyware and malware within the Google Play Store, in February 2012, Google introduced the Bouncer service, which tests applications in order to identify presence of any Spyware, Trojans, Malware and other suspected behaviors (Ziegler 2012). However, Google has failed to provide specific information about how exactly Bouncer works and how successful it truly is. On the other hand, several studies executed by Oberheide (Oberheide 2012) demonstrated that, when subjected to dynamic runtime examination of Android applications in a virtual emulation environment, Bouncer can be quickly bypassed. In conclusion, the author noted that Bouncer focused on finding malicious applications, rather than privacy-intrusive ones.

Numerous studies have been conducted in an attempt to provide methods and instruments capable of identifying sensitive data leakage within Smartphone programs (Enck et al. 2009; Barrera et al. 2010; Kane 2010; Beresford et al. 2011; Chin et al. 2011; Enck 2011; Enck et al. 2011; Felt et al. 2011; Felt et al. 2011; Hornyack et al. 2011; Vidas et al. 2011; Zhou et al. 2011; Enck et al. 2014; Store 2014). While there are several approaches to this issue, three procedures tend to be predominantly used in application investigation—static code, dynamic flow, and permission analysis. While extant research in this field is categorized in Table 7, Table 8 displays studies based on the specific methods researchers used. Presently, a large number of security and privacy extensions has been suggested, most of which aimed to offer users much more control of resources within their mobile phones (Nauman et al. 2010; Beresford et al. 2011; Jeon et al. 2011; Zhou et al. 2011; Pearce et al. 2012). In the following sections, these functions will be explored in depth.

Table 7: Changes and improvements with respect to the new framework design

| Features in Privacy Implementation | APEX | Kiran | SAINT | Mockdroid | XmanDroid | TISSA | aSpotCat | Permission Dog | Permission Denied | LBE privacy guard | Cyanogen Mod | WhisperCore |
|---|---|---|---|---|---|---|---|---|---|---|---|---|
| 1-Availability | - | - | - | CR | - | - | A | A | A | A | CR | CR |
| 2-Modified OS | √ | √ | √ | √ | √ | √ | - | - | - | - | √ | √ |
| 3-Policy | - | √ | √ | - | √ | - | - | - | - | - | - | - |
| 4-Conditions | √ | - | - | - | - | - | - | - | - | - | - | - |
| 5-Blocking | √ | - | √ | - | √ | - | - | - | √ | √ | √ | - |
| 6-Pseudo technique | - | - | - | √ | - | √ | - | - | - | - | - | √ |
| 7-User Confirmation | - | - | - | - | - | - | - | - | - | √ | - | - |
| 8-Cloud support | - | - | - | - | - | - | - | - | - | - | - | - |
| 9-Log | - | - | - | - | - | - | - | - | - | √ | - | - |
| 10-GUI | - | - | - | - | - | - | √ | √ | √ | √ | - | - |

Table 8: Studies based on methods used

| | Static Analysis | Dynamic Analysis | Permission Analysis |
|---|---|---|---|
| **Researches Example** | (Chin et al. 2011) | (Kane 2010) | (Enck et al. 2009) |
| | (Felt et al. 2011) | (Enck et al. 2014) | (Barrera et al. 2010) |
| | (Felt et al. 2012) | (Beresford et al. 2011) | (Felt et al. 2011) |
| | (Enck et al. 2011) | (Zhou et al. 2011) | (Felt et al. 2011) |
| | (Store 2014) | (Hornyack et al. 2011) | (Vidas et al. 2011) |
| | App Profiles [18] | (Yang et al. 2012) | (Book et al. 2013) |
| | | | (Frank et al. 2012) |

After an extensive study of different types of analysis, static code, dynamic flow, and permission analysis have been noted to be most commonly used and are thus explored in detail in the subsequent sections.

- **Static Analysis**

This kind of program examination could be carried out with or without the source code. Thus far, many mobile applications have been subjected to static analysis, typically requiring decompilers to recover the application source code, such as (Code.google.com 2014). Among 1400 applications analyzed, over 50% were found to pose potential privacy threat to the system ID, without the mobile user being aware of this issue. Based on this review, Chin et al. (2011) proposed ComDroid, claiming that it could provide user protection by utilizing disassembled DEX bytecode. More specifically, ComDroid recognizes vulnerabilities within the intent communications among apps, such as broadcast theft, assistance hijacking, and many others (Chin et al. 2011). Among 100 applications studied, the authors observed 34 inconveniences for the user. In another study of this type, Application Profiles (Store 2014) created by research group called RobustNet at the Umich (Qiang (Chad) Xu 2014) were used to examine cell phone apps offline, with the goal of identifying privacy-related behaviors included to the application program.

As these studies show, although static evaluation offers a comprehensive computerized search and scan of Smartphone applications, reliability of its findings is strongly dependent on the functionality on the decompiler utilized or the code design the specific programmer adopted. For that reason, its performance cannot be guaranteed and will always be context-specific. A further difficulty pertains to the technique's inability to distinguish between real and fake privacy-related actions from the Smartphone owner's perspective.

- **Dynamic Analysis**

Dynamic examination might help assist in mitigating the current ambiguity pertaining to authorization granularity and also offer a user-friendly approach for monitoring the manner in which individual applications are executed. The Wall Street Journal reviewed and analyzed 101 favorite mobile apps by keeping track of network communications (Kane 2010). The findings revealed that of these 101 apps, 56 transported the actual Smartphone's unique ID to a third party hosting service without seeking explicit user permission. In addition, 47 application programs shared not only the Smartphone's actual location, but also transmitted other sensitive data, including user's gender, age group, etc., without the user's knowledge.

In a recent study, TaintDroid conducted a comprehensive dynamic flow investigation in order to assess the extent of data leakage on Android Smartphone devices (Enck et al. 2014). Additional jobs were performed in TaintDroid, with the aim of conducting further privacy studies and potentially proposing useful adjustments to the OS (Beresford et al. 2011; Hornyack et al. 2011). An earlier study conducted by Yang et al. (2012) employed crowdsourcing in a dynamic research framework to highlight the need for application-specific permissions.

Dynamic investigation and analysis is useful, as it provides information on what exactly occurs while a software application is operating. However, its main disadvantage is its restricted scalability, since human interventions (communications along wearable and portable applications) are essential in order to prompt a particular software application behaviors that are needed for analysis and investigation.

While this approach is certainly more useful than static analysis, it cannot easily distinguish user preferences in terms of privacy settings and focus on applications that can potentially act maliciously once installed on user's Smartphone. In other words, it does not approach cell phone privacy from the users' viewpoint, which this study will aim to address through the use of cloud-based smart decision-making system. The goal is to be able to connect the actual space concerning application evaluation while accounting for the main privacy considerations most users expect. For this purpose, this study utilizes permission replacement and obtains a new set of permissions that may help increase user privacy on Smartphone devices.

- **Permission Analysis**

Through examining the permission lists reported through the software application program, potentially high-risk functionalities might be recognized. This particular type of research has yielded some promising results (Barrera et al. 2010; Felt et al. 2011; Felt et al. 2011; Vidas et al. 2011). The reported findings illustrate frequent consumption patterns (Barrera et al. 2010), misuses (Felt et al. 2011; Vidas et al. 2011), as well as potential implications for security and privacy in the Android operating system (Felt et al. 2011; Felt et al. 2011; Felt et al. 2011). The study conducted by Enck et al. (2009) was among the first to incorporate permission research and analysis in the assessment of the Android operating system.

Barrera et al. (2010) also conducted a permission investigation of 1100 cost-free software applications within the Google Play Store repertoire. They reported rapid decay in the number of software applications that requested particular permissions from the user, as nearly all applications examined in this work could operate with limited permissions (Barrera et al. 2010). The following year, Felt et al. (2011) analyzed the potency of permissions within the setup time. The goal was to identify application designers that made mistakes when declaring application permissions list requirements (by, for example, asking for unnecessary or non-useful permissions). In order to mitigate this issue, Felt et al. proposed a new approach, referred to as Stowaway, which executed static evaluation in order to help identify over-privileged software applications. In the same manner, SDK extension for Android was created by Vidas et al. (Vidas et al. 2011), with the goal of helping Android application developers identify and publish the minimum amount of permissions needed for a particular app's performance.

In their study, Frank et al. (2012) explored permission seeking behaviors of various Android software applications. By applying matrix factorization methods, the authors discovered 30 regular permission seeking patterns. In a more recent work by Book et al. (2013), the authors assessed 144000 Smartphone applications in order to compile a comprehensive library consumption list of permissions (Book et al. 2013). The outcome was an assessment that revealed which advertisement libraries' utilization of permissions has considerably increased during the last few years.

As was shown by the aforementioned studies, utilization of sensitive information by the applications creates unique challenges to consumer privacy and security. Clearly, being able to grant application-specific permissions is important, as it not only enhances device performance but also increases user safety. However, as indicated above, permission lists cannot offer detailed information regarding the private resources that are utilized and for what purpose. As a result, these approaches offer limited security and privacy to Smartphone users.

## 4. CONCLUSIONS AND FUTURE WORK

In this paper, a different perspective on the Android Smartphone privacy problem was provided. Instead of focusing on application-based actions, the researcher designed a framework that can benefit non-technical users, by enabling them to control their privacy automatically or manually. This was achieved by controlling permissions requested by every application installed on the mobile device and using a cloud decision-making system to provide new set of permissions for every application.

The CenterYou framework allows advanced users to manage their own app permissions. While it is expected that most users will choose the Autopilot mode, they can easily switch from one to another, if needed. This flexibility, ease of use and robustness of CenterYou is expected to be highly beneficial to the growing body of Android users. The simple cloud-based decision-making system can protect non-technical users from privacy risks, while also providing them with confidence that their data is protected, without the need for any input on their part. Above all, CenterYou dispenses with having to choose permissions for every application users wish to install on their device. With the help of the CenterYou application, private data will stay private, while users can benefit from new apps that are emerging daily on the market. In addition, CenterYou provides protection for both corporate and individual clients. The application is designed to provide benefits to organizations as well as individuals, and this is likely one of the most important parts of the CenterYou innovation.

Because this solution likely requires root access on the device, there should be alternative ways to control the privileges requested by applications. Thus, future work in this field could focus on the following objectives:

- Implement a service to separate the applications by categorizing them. This can help users to keep the best control on every category.
- Design an isolated service to separate the applications used for work from those that are suitable only for home usage. This can be achieved by recording these groups of apps in a different environment that can be moderated by the user or cloud services, depending on the mode of operation the user has selected.

## ACKNOWLEDGEMENTS

The CenterYou solution design is registered for intellectual property (File Number: UKM3.2.29/108/2/718) of UKM, Malaysia. We are very thankful to anonymous reviewers for their comments, reply and suggestions for CenterYou: A cloud-based Approach to Simplify Android Privacy Management, which helps and improve the future researchers.